

\baselineskip=15pt
\magnification=1200
\hsize 15.5truecm\hoffset 1.2truecm
\vsize 22.0truecm\voffset 1.5truecm
\outer\def\beginsection#1\par{\medbreak\bigskip
      \message{#1}\centerline{\bf#1}\nobreak\medskip\vskip-\parskip
      \noindent}
\font\grande=cmbx10 scaled\magstep1

\def \La {\Lambda}
\def \Da {\Delta}
\def \a {\alpha}
\def \Ga {\Gamma}

\def \sg {\sigma}
\def \da {\delta}

\def \th {\theta}
\def \Sg {\Sigma}
\def \vphi {\varphi}
\def \noi {\noindent}
\def \rline {\rightline}
\def\sitontop#1#2{\mathrel{\mathop{#1}\limits^{\scriptscriptstyle #2}}}
\def\tr{{\rm Tr}}
\def\um{{1\over2}}
\def\o#1#2{{#1\over#2}}

\def\pa{\partial}

\rline {DFTT 54/93}
\rline {gr-qc/9311029}
\rline {October 1993}
\vskip 1truecm

\centerline{\grande Quantisation of 2+1 gravity for genus 2}
\vskip 1truecm
\centerline {J.E.Nelson and T.Regge}
\vskip1truecm
\centerline {Dipartimento di Fisica Teorica dell'Universit\`a - Torino
-Italy}
\centerline {Via Pietro Giuria 1, 10125,Torino, Italy}
\centerline{\it email:~ nelson@to.infn.it,~ telefax ~039-11-6707214}
\vskip 1truecm

In [1,2] we established and discussed the algebra of observables
for $2+1$ gravity at both the classical and quantum level, and gave a
systematic discussion of the reduction of the expected number of
independent observables to $6g - 6 (g > 1)$. In this paper the algebra
of observables for the case $g=2$ is reduced to a very simple form. A
Hilbert space of state vectors is defined and its representations are
discussed using a deformation of the Euler-Gamma function . The
deformation parameter $\th$ depends on the cosmological and Planck's
constants.\par
\noi P.A.C.S. 04.60

\beginsection 1. Introduction.

In a previous article [1] we presented the abstract quantum
algebra for $2+1$ gravity with cosmological constant $\La$:

$$
\eqalignno{
(a_{mk},a_{jl})\ &=\ (a_{mj},a_{kl})\ =\ 0 & (1.1) \cr
(a_{jk},a_{km})\ &=\ \Big(\o {1} {K}-1\Big)(a_{jm}-a_{jk}a_{km}) & (1.2)
\cr
(a_{jk},a_{kl})\ &=\ \Big(1-\o {1} {K}\Big)(a_{jl}-a_{kl}a_{jk}) & (1.3)
\cr
(a_{jk},a_{lm})\ &=\ \Big(K-\o {1} {K}\Big)(a_{jl}a_{km}-a_{kl}a_{jm}) &
 (1.4) \cr}
$$
where $K\ =\ \o {4\a -ih} {4\a +ih}\ =\ e^{i\th},\ \La\ =\ -\o {1}
{3\a^{2}}$ is the cosmological constant and h is Planck's constant .
In  (1.1-4) $m,j,l,k$. are 4 anticlockwise points of Fig.1. $m,j,l,k\ =\
1\cdots n$, and the time independent quantum operators $a_{lk}$
correspond to the classical $\o {n(n-1)} {2}$ gauge invariant trace
elements

$$
\a_{ij}\ =\ \a_{ji}\ =\ \um\tr\Big(S(t_{i}t_{i+1}\cdots t_{j-1})\Big)\ ,
\ S\in SL(2,R)
\eqno (1.5)
$$
For $n\ =\ 2g+2$ the map $S\ :\ \pi_{1}(\Sg)\sitontop {\longrightarrow}
{S} SL(2,R)$ is defined by the integrated anti-De~Sitter connection in
the initial data Riemann surface $\Sg$ of genus $g$, and refers to one
of the two spinor components, say the upper component, of the spinor
group $SL(2,R)\otimes SL(2,R)$ of the gauge group $SO(2,2)$ of $2+1$
gravity with negative cosmological constant [2]. The algebra (1.1-4) is
invariant under the quantum action of the mapping class group on traces
[1], the lower component yields an independent algebra of traces
$b_{ij}$ identical to (1.1-4) but with $K\to 1/K$. Moreover $(a_{ij},
b_{kl})\ =\ 0\ \forall\ i,j,k,l$. Here we discuss only the upper
component. The homotopy group $\pi_{1}(\Sg)$ of the surface is defined
by generators $t_{i},\ i\ =\ 1\cdots 2g+2$ and presentation:

$$
t_{1}t_{2}\cdots t_{2g+2}\ =\ 1\ ,\ t_{1}t_{3}\cdots t_{2g+1}\ =\ 1\ ,\
t_{2}t_{4}\cdots t_{2g+2}\ =\ 1 \eqno (1.6)
$$
The first relator in (1.6) implies that $\Sg$ is closed. The operators
in (1.1-4) are ordered with the convention that $t(a_{ij})$ is
increasing from left to right where $t(a_{ij})\ =\ \o {(i-1)(2n-2-i)} {2}
+j-1$.

The case of $g = 1$, the torus, has been studied extensively, both in
this approach [2], and others [3,4]. In this approach the algebra
(1.1-4) is isomorphic to the quantum algebra of $SU(2)_{q}$ when
$\La\not= 0$ [2]. For $\La =0$ it has been shown [5] that the metric
approach to determining the complex modulus of the torus [3] is
classically equivalent to the classical limit of (1.1-4) for $n = 4$.
There are similar, recent results for $\La\not= 0$ [6].

For $g > 1$ there are very few results apart from those of Moncrief [3]
who studies the second order, metric formalism and achieves very
general results. In this article the case $g = 2$ of the algebra (1.1-4)
is studied in detail. In [7] we determined for $n\leq 6$, i.e. $g\leq 2$
a set of $p$ linearly independent central elements $A_{nm},\ m\ =\ 1
\cdots p$ where $n\ =\ 2p$ or $n\ =\ 2p+1$, and analysed the trace
identities which follow from the presentation (1.5) of the homotopy
group $\pi_{1}(\Sg)$ and a set of rank identities. These identities
together generate a two-sided ideal. For generic $g$ there are
precisely $6g-6$ independent elements which satisfy the algebra (1.1-4).
The reduction from $\o {n(n-1)} {2}\ =\ (g+1)(2g+1)$ to $6g-6$ results
from the use of the above mentioned identities [7] but is highly non
unique. For $g=2$ the reduction from the original 15 elements $a_{ij}$
to 6 independent elements has been the subject of a long study. Here
this reduction is implemented explicitly in terms of a set of 6
independent operators which satisfy a particularly simple algebra. There
are many such possibilities but a convincing set is described as follows:

We group the vertices of the hexagon into 3 sectors, see Fig.2, the
vertices labeled $2b$ and $2b-1$ belonging to the sector
$b,\ b\ =\ 1\cdots 3$. Accordingly we define the sector function $s[2b]\
=\ s[2b-1]\ =\ b$. A convenient choice for the 6 independent elements
is given by 3 commuting angles $\vphi_{-b}\ =\ -\vphi_{b},\ b\ =\ \pm 1
\cdots\pm 3$ defined by:

$$
a_{2b-1,2b}\ =\ \o {\cos\vphi_{b}} {\cos\o {\th} {2}}\qquad b\ =\ 1
\cdots 3 \eqno (1.7)
$$
and commuting operators $M_{ab}$ with the properties:

$$
\eqalign{
M_{ab}\ &=\ M_{ba}\qquad a,b\ =\ \pm 1\cdots\pm 3 \cr
M_{a,-a}\ &=\ 1,\ M_{a,-b}M_{b,c}\ =\ M_{ac} \cr} \eqno (1.8)
$$
The $M_{ab}$ act as raising and lowering operators on the $\vphi_{a}$:

$$
M_{\pm a,b}\vphi_{a}\ =\ (\vphi_{a}\mp\th)M_{\pm a,b} \eqno (1.9)
$$
It can be checked that the 12 remaining $a_{ik}$ are represented by:

$$
\eqalign{
a_{kj}\ &=\ \o {1} {K+1} K^{\o {\tilde k +\tilde j} {2}+1}
\sum_{n,m=\pm 1}\ \exp(-i(n(\tilde k +1)\vphi_{a}+m\tilde j \vphi_{b}))
\times \cr
&\ \times\o {\sin\left(\o {\th} {4}+\o {n\vphi_{a}+m\vphi_{b}+\vphi_{c}}
{2}\right)\sin\left(\o {\th} {4}+\o {n\vphi_{a}+m\vphi_{b}-\vphi_{c}}
{2}\right)} {\sin(n\vphi_{a})\ \sin(m\vphi_{b})} M_{na,mb} \cr}
\eqno (1.10)
$$
where we set $\tilde k \ =\ k\ {\rm mod}\ 2+\um$ and $a\ =\ s(k),\ b =
s(j),\ a,b,c$ in cyclical order.\par
\noi Under these conditions the $a_{ik}$ satisfy the trace and rank
identities. These identities can all be derived from:

$$
a_{12}a_{34}+K^{-2}a_{23}a_{14}-K^{-1}a_{13}a_{24}-a_{56}\ =\ 0
\eqno (1.11)
$$
by repeated commutation with the elements of the algebra (1.1-4). For
example two useful identities are:

$$
K^{3}a_{12}a_{46}+Ka_{24}a_{16}-K(1+K^{3})a_{34}a_{45}-K^{2}a_{14}a_{26}
+(1-K+K^{2})a_{35}\ =\ 0 \eqno (1.12)
$$

$$
\eqalign{
&(1+K^{3})((1+K)a_{34}a_{56}a_{45}-Ka_{34}a_{46}-a_{56}a_{35})+\cr
+K^{2}a_{14}a_{25}&-K^{3}a_{12}a_{45}-Ka_{24}a_{15}+K(1+K^{2}-K)
a_{36}\ =\ 0 \cr}
$$
and their images under cyclical permutations of the indices $1\cdots 6$.
These identities are certainly not all independent. By heavy use of
computer algebra we were able to show that (1.11-12) and their images
follow from (1.7-10).\par
\noi The relations (1.8-9) follows from the single sector factorisation
for all $a,b\ =\ \pm 1\cdots \pm 3$:

$$
M_{ab}\ =\ M_{a}M_{b}\ =\ M_{ba}\ =\ M_{b}M_{a} \eqno (1.13)
$$

$$
M_{-a}\ =\ M_{a}^{-1} \eqno (1.14)
$$

$$
M_{\pm a}\vphi_{a}\ =\ (\vphi_{a}\mp\th)M_{\pm a} \eqno (1.15)
$$
It is clear that (1.13-15) can be formally satisfied by setting
$M_{a}\ =\ \exp\Big(-\th\o {\pa} {\pa\vphi_{a}}\Big)$ and therefore
$M_{ab}\ =\ \exp\Big( -\th\big(\o {\pa} {\pa\vphi_{a}}+\o {\pa}
{\pa\vphi_{b}}\big)\Big)$, in turn (1.10) becomes:

$$
\eqalign{
a_{kj}\ &=\ \o {1} {2\cos\left(\o {\th} {2}\right)}\sum_{n,m=\pm 1} \o
{\sin\left(\o {\th} {4}+\o {n\vphi_{a}+m\vphi_{b}+\vphi_{c}} {2}\right)
\sin\left(\o {\th} {4}+\o {n\vphi_{a}+m\vphi_{b}-\vphi_{c}} {2}\right)}
{\sin n\vphi_{a}\ \sin m\vphi_{b}} \times \cr
&\ \times\ \exp\Big(-i\big(n(\tilde k
+1)\vphi_{a}+m\tilde j\vphi_{b}+\th(np_{a}+mp_{b})\big)\Big) \cr}
\eqno (1.16)
$$
where we have used the Baker-Hausdorff formula [8]:

$$
\exp A\ \exp B\ =\ \exp \left(A+B+\o {AB-BA} {2}\right)\ =\ \exp (AB-BA)
\exp B\ \exp A \eqno (1.17)
$$
valid when $AB-BA$ is a $C$-number and $M_{a}\ =\ \exp(-i\th p_{a})$.
Note that, from (1.7) and (1.16), all of the 15 original $a_{ij}$ are
expressed in terms of the 3 angles $\vphi_{a}$ and their conjugate
momenta $p_{a}$.\par
\noi The treatment of (1.16) can be further simplified by noting that
$a_{kj}\ =\ U_{\tilde k\tilde j}^{-1}A_{ab}U_{\tilde k\tilde j}$ where:

$$
U_{\tilde k\tilde j}\ =\ \exp\left(i\o {(\tilde k+1)\vphi_{a}^{2}+
\tilde j\vphi_{b}^{2}} {2\th}\right) \eqno (1.18)
$$

$$
\eqalign{
A_{ab}\ &=\ \o {1} {2\cos\left(\o {\th} {2}\right)}\sum_{n,m=\pm 1} \o
{\sin\left(\o {\th} {4}+\o {n\vphi_{a}+m\vphi_{b}+\vphi_{c}} {2}\right)
\sin\left(\o {\th} {4}+\o {n\vphi_{a}+m\vphi_{b}-\vphi_{c}} {2}\right)}
{\sin n\vphi_{a}\ \sin m\vphi_{b}}\times \cr
&\ \times\exp\big(-i\th(np_{a}+mp_{b})\big) \cr}
$$
$A_{ab}$ is an operator which is a function of the sectors $a,b$ only
and is independent of the position of $k,j$ within $a,b$.

The discussion of the representations of (1.13-15) is considerably
simplified by the introduction of the deformed Euler Gamma-function $\Ga
(z,\th)$ (see Appendix for the definition and a list of properties)
which extends to the complex domain the symbol:

$$
\Big[n\Big]!\ =\ \prod_{p=1}^{n}\o {\sin\o {p\th} {2}} {\sin\o {\th}
{2}} \eqno (1.19)
$$
In particular $\Ga(n+1,\th)\ =\ \Big[n\Big]!$ and $\Ga(z,\th)$ is a
meromorphic analytic function of $z$ with poles at $z\ =\ -s-\o {2\pi r}
{\th},\quad s,r\geq 0$ and integer and zeroes at $z=s+\o {2\pi r}
{\th}$, $s,r\geq 1$ and integer.

\beginsection 2. Representations.

The $a_{ij}$ expressed by (1.7) are by definition all hermitian
operators. We denote by $\phi_{a}$ the generic eigenvalue of the
operator $\vphi_{a}$ and set $\phi\ =\ \{\phi_{1},\phi_{2},\phi_{3}\},\
z_{a}\ =\ \cos\phi_{a},\ z\ =\ \{z_{1},z_{2},z_{3}\}$ where the $z_{a}$
are real and restricted to a domain $D^{3}\subset R^{3}$. Let $T$ with
$T^{2}=1$ be the antilinear conjugacy operator $\Psi(z)\sitontop
{\longrightarrow} {T} \Psi^{*}(z)$. A measure $\sg(z)d^{3}z$ with
$\sg(z)\geq 0$ and real turns $H$ into a Hilbert space $H$ with norm:

$$
\vert\Psi\vert^{2}\ =\ \int\limits_{D^{3}}\vert\Psi(z)\vert^{2}\sg(z)
d^{3}z \eqno (2.1)
$$
The weight function $\sg(z)$ can be determined from the hermiticity of
the $a_{ij}$ (1.7) as follows.\par
\noi Let $p_{a}\ =\ -i\o {\pa} {\pa\phi_{a}},\quad a=1,2,3$
satisfying the CCR:

$$
(\vphi_{a},\vphi_{b})\ =\ 0,\ (\vphi_{a},p_{b})\ =\ i\da_{ab},\ (p_{a},
p_{b})\ =\ 0,\ a,b,\ =\ 1,2,3 \eqno (2.2)
$$
it follows by conjugation that:

$$
(\vphi_{a}^{\dag},\vphi_{b}^{\dag})\ =\ 0,\ (\vphi_{a}^{\dag},
p_{b}^{\dag})\ =\ i\da_{ab},\ (p_{a}^{\dag},p_{b}^{\dag})\
=\ 0,\ a,b,\ =\ 1,2,3 \eqno (2.3)
$$
but also that

$$
\vphi_{a}^{\dag}\ =\ T\vphi_{a}T,\ a_{2a,2a-1}\ =\ Ta_{2a,2a-1}T,\
a\ =\ 1,2,3 \eqno (2.4)
$$
The hermiticity relation between $O,O^{\dag}$ namely
$\langle\Psi,O\Phi\rangle^{*}\ =\ \langle O^{\dag}\Psi,\Phi\rangle$
implies $\left(-i\o {\pa} {\pa z_{a}}\right)^{\dag}\
=\ -i\sg^{-1}\o {\pa} {\pa
z_{a}}\sg$. But $\o {\pa} {\pa z_{a}}\ =\ \o {-1} {\sin\phi_{a}}\o
{\pa} {\pa\phi_{a}}$ whereby:

$$
\eqalign{
p_{a}^{\dag}\ &=\ \left(-i\o {\pa} {\pa\phi_{a}}\right)^{\dag} \ =\
\left(
i\sin\phi_{a}\o {\pa} {\pa z_{a}}\right)^{\dag}\ =\ i\sg^{-1}\o {\pa}
{\pa z_{a}}\sg\ \sin\phi_{a}^{+}\ =\ \cr
\ &=\ i\sg^{-1}\o {\pa} {\pa z_{a}}\sg T\sin\phi_{a}T\ =\ -iT
\sg^{-1}\o {\pa} {\pa z_{a}}\sg\ \sin\phi_{a}T\ =\ -T\rho^{-1}
p_{a}\rho T \cr } \eqno (2.5)
$$
where $\rho(\phi)\ =\ C\sin\phi_{1}\sin\phi_{2}\sin\phi_{3}\sg(z),\ C$
being a normalization constant, the operator $\rho\ =\ \rho(\vphi_{1},
\vphi_{2},\vphi_{3})\ =\ \rho(\vphi)$ is now to be determined by
extending (2.4) to all $i,k$ as $a_{ik}\ =\ a_{ik}^{\dag}\ =\
Ta_{ik}T$.\par
\noi From (1.16-18) we obtain by conjugation:

$$
\eqalign{
a_{kj}\ &=\ \o {1} {2\cos\left(\o {\th} {2}\right)}U_{\tilde k
\tilde j}^{\dag}\sum_{n,m=\pm 1} \exp \Big( i\th(np_{a}^{\dag}
+mp_{b}^{\dag})\Big)\times\cr
&\ \times\o {\sin\left(\o {\th} {4}+\o {n\vphi_{a}^{\dag}+m
\vphi_{b}^{\dag}
+\vphi_{c}^{\dag}} {2}\right)\sin\left(\o {\th} {4}+\o {n\vphi_{a}^{\dag}
+m\vphi_{b}^{\dag}-\vphi_{c}^{\dag}} {2}\right)} {\sin n\vphi_{a}^{\dag}
\sin m\vphi_{b}^{\dag}}\ U_{\tilde k\tilde j}^{-1,\dag}\cr} \eqno (2.6)
$$
We apply now [1.17] and reorder the operators in (2.6) by bringing the
exponential factor to the right thus finding:

$$
\eqalign{
a_{kj}\ &=\ \o {1} {2\cos\left(\o {\th} {2}\right)}U_{\tilde k
\tilde j}^{\dag}\sum_{n,m=\pm 1} \o {\sin\left(\o {5\th} {4}+\o {n
\vphi_{a}^{\dag}+m\vphi_{b}^{\dag}+\vphi_{c}^{\dag}} {2}\right)\sin
\left(\o {5\th} {4}+\o {n\vphi_{a}^{\dag}+m\vphi_{b}^{\dag}
-\vphi_{c}^{\dag}} {2}
\right)} {\sin (n\vphi_{a}^{\dag}+\th)\ \sin (m\vphi_{b}^{\dag}+\th)}
\times\cr
&\ \times\exp \Big( i\th(np_{a}^{\dag}+mp_{b}^{\dag})\Big) U_{\tilde k
\tilde j}^{-1,\dag}\cr}
$$
and

$$
\eqalign{
a_{kj}\ &=\ \o {1} {2\cos\left(\o {\th} {2}\right)}TU_{\tilde k
\tilde j}^{-1}\sum_{n,m=\pm 1} \o {\sin\left(\o {5\th} {4}+\o {n
\vphi_{a}+m\vphi_{b}+\vphi_{c}} {2}\right)\sin\left(\o {5\th} {4}
+\o {n\vphi_{a}+m\vphi_{b}-\vphi_{c}} {2}\right)} {\sin (n\vphi_{a}+\th)
\ \sin (m\vphi_{b}+\th)}\times\cr
&\ \times\exp \Big( i\th(np_{a}+mp_{b})\Big) U_{\tilde k\tilde j}T \cr}
\eqno (2.7)
$$
We define the maps:

$$
\phi_{a},\phi_{b},\phi_{c}\sitontop {\longrightarrow} {\Da(na,mb)}
\phi_{a}+n\th,\phi_{b}+m\th,\phi_{c} \eqno (2.8)
$$
where as before $n,m$ take all values $\pm 1$ and $a,b,c$ are any
permutation of $1,2,3$. From $a_{ik}\ =\ a_{ik}^{\dag}\ =\ Ta_{ik}T$
and by
comparing (2.7) with (1.16) we find the recursion relation in the
eigenvalues $\phi,z$:

$$
\eqalign{
\Da(na,mb)\sg(z_{1},z_{2},z_{3})\ &=\ \sg(z_{1},z_{2},z_{3})\times\cr
&\ \times\o {\sin\left(\o {\th} {4}-\o {n\vphi_{a}+m\vphi_{b}+\vphi_{c}}
{2}\right)\sin\left(\o {\th} {4}-\o {n\vphi_{a}+m\vphi_{b}-\vphi_{c}}
{2}\right)}
{\sin\left(\o {5\th} {4}+\o {n\vphi_{a}+m\vphi_{b}+\vphi_{c}} {2}\right)
\sin\left(\o {5\th} {4}+\o {n\vphi_{a}+m\vphi_{b}-\vphi_{c}} {2}\right)}
\cr} \eqno (2.9)
$$
A solution of (2.9) is  then provided by:

$$
\sg(z_{1},z_{2},z_{3})\ =\ P(\phi)\prod_{m_{1}m_{2}m_{3}=\pm 1}\Ga\left(
-\o {1} {4}+\o {q\pi} {\th}+\o {m_{1}\phi_{1}+m_{2}\phi_{2}+m_{3}
\phi_{3}} {2\th},2\th\right) \eqno (2.10)
$$
where $q$ is arbitrary and integer and $P(\phi)$ is invariant under
(2.8), otherwise arbitrary, in (2.10) the product is carried on all
independent sign choices of $m_{1},m_{2},m_{3}$.\par
\noi By using (A.7) we see that:

$$
E(\phi,\th,q+1)\ =\ S(\phi)E(\phi,\th,q)
$$
where:

$$
S(\phi)\ =\ 2^{8}(2\sin\th)^{-\o {8\pi} {\th}}
\prod_{m_{1},m_{2},m_{3}=\pm 1}\sin\pi\left( -\o {1} {4}+\o {q\pi} {\th}
+\o {m_{1}\phi_{1}+m_{2}\phi_{2}+m_{3}\phi_{3}} {2\th}\right)
$$

$$
E(\phi,\th,q)\ =\ \prod_{m_{1},m_{2},m_{3}=\pm 1}\Ga\left( -\o {1} {4}
+\o {q\pi} {\th}+\o {m_{1}\phi_{1}+m_{2}\phi_{2}+m_{3}\phi_{3}} {2\th},
2\th\right) \eqno (2.11)
$$
Since $S(\phi)$ is invariant under (2.8) it can be absorbed into
$P(\vphi)$ hence the appearance of $q$ does not signal any new
arbitrariness. It is however convenient in our discussion to have a
solution which depends explicitly on $q$.\par
\noi The function $\rho(\phi)$ is periodic of period $2\pi$ and odd in
$\phi_{1},\phi_{2},\phi_{3}$ if we have (see (2.10) and (A.7)):

$$
\eqalign{
\o {\rho(\phi_{1}+2\pi,\phi_{2},\phi_{3})} {\rho(\phi_{1},\phi_{2},
\phi_{3})}\ &=\ \o {P(\phi_{1}+2\pi,\phi_{2},\phi_{3})} {P(\phi_{1},
\phi_{2},\phi_{3})}\times\cr
&\ \times\prod_{m_{2}m_{3}=\pm 1} \o {\sin\pi\left( -\o {1}
{4}+\o {q\pi} {\th}+\o {\phi_{1}+m_{2}\phi_{2}+m_{3}\phi_{3}}
{2\th}\right)} {\sin\pi\left(\o {1} {4}-\o {(q-1)\pi} {\th}+\o {\phi_{1}
+m_{2}\phi_{2}+m_{3}\phi_{3}} {2\th}\right)}\ =\ 1 \cr}
$$
This be achieved by setting

$$
\th\ =\ \o {2q-1} {2t+1}2\pi\quad,\quad t\ {\rm integer} \eqno (2.12)
$$
and $P(\vphi)\ =\ 1$.\par
\noi We list here the basic properties of $E(\phi,\th,q)$:

\item {1).} $E(\phi,\th,q)$ is even in each of the $\phi_{1},\phi_{2},
\phi_{3}$.

\item {2).} $E(\phi,\th,q)$ is periodic of period $2\pi$ in each of the
$\phi_{1},\phi_{2},\phi_{3}$.

\item {3).}$E(\phi,\th,q)$ is real but not necessarily positive for
$\phi_{1},\phi_{2},\phi_{3}$ all real. It follows by analytic
continuation that $E(\phi^{\dag},\th,q)\ =\ E(\phi,\th,q)^{\dag}$.

\item {4).} $E(\phi,\th,q)$ is real and positive if at least one of the
$\phi_{1},\phi_{2},\phi_{3}$ is imaginary and the others real. This
follows from the possibility of arranging (2.11) in pairs of conjugate
factors.\par
\noi In this case we may choose $\sg(z)\ =\ E(\phi,\th,q)$. The
discussion of the positivity of the function $\sg(z)$ for arbitrary $z$
is rather involved. A particular solution is provided by restricting all
$z_{a}$ to the hyperbolic domain $z_{a}>1$, i.e. all $\phi_{a}$ pure
imaginary. In this case all $a_{kj}$ from (1.7) and (1.16) are
represented by unbounded hermitian operators. This, and the inclusion
of the other $SL(2,R)$ component, will be discussed elsewhere [9].

\beginsection Appendix.

Here we give the definition and a comprehensive list of properties of
the deformed Euler Gamma function:

$$
\Ga(z,\th)\ =\ \left(\o {\th} {2\sin\o {\th} {2}}\right)^{z-1}\Ga(z)
\prod_{n=1}^{\infty}\left(\o {\th} {2\pi n}\right)^{2z-1}\o {\Ga\left(z+
\o {2\pi n} {\th}\right)} {\Ga\left(1-z+\o {2\pi n} {\th}\right)}
\eqno (A.1)
$$

$$
\lim_{\th\to 0}\Ga(z,\th)\ =\ \Ga(z) \eqno (A.2)
$$

$$
\Ga(1,\th)\ =\ 1 \eqno (A.3)
$$

$$
\Ga(z+1,\th)\ =\ \Ga(z,\th)\o {\sin\o {\th z} {2}} {\sin\o {\th} {2}},\
\Ga(n+1,\th)\ =\ \Big[n\Big]!\ n\ {\rm integer}\ >0 \eqno (A.4)
$$
$$
\Ga\left(z+\o {2\pi} {\th},\th\right)\ =\ 2\sin(\pi z)\left(2\sin
\o {\th} {2}\right)^{-\o {2\pi} {\th}}\Ga(z,\th) \eqno (A.5)
$$

$$
\Ga(z,\th)\Ga(1-z,\th)\ =\ \o {2\pi\sin\o {\th} {2}} {\th\sin(\pi z)}
\eqno (A.6)
$$

$$
\Ga(z,\th)\Ga\left(\o {2\pi} {\th}-z,\th\right)\ =\ \o {\pi} {\th\sin\o
{\th z} {2}}\left(2\sin\o {\th} {2}\right)^{2-\o {2\pi} {\th}}
\eqno (A.7)
$$

$$
\Ga(z,\th)\Ga\left(1+\o {2\pi} {\th}-z,\th\right)\ =\ \o {2\pi}
{\th}\left(2\sin\o {\th} {2}\right)^{1-\o {2\pi} {\th}} \eqno (A.8)
$$
Setting $\th^{\prime}\ =\ \o {4\pi^{2}} {\th}$ we have the duality
property:

$$
\Ga(z.\th)\ =\ \Ga\left(\o {\th z} {2\pi},\th^{\prime}\right)\ \left(2
\sin\o {\th^{\prime}} {2}\right)^{\o {z\th} {2\pi}-1}\left(2\sin\o {\th}
{2}\right)^{1-z}\o {\th^{\prime}} {2\pi} \eqno (A.9)
$$
(A.9) is meaningless in the limit $\th\to 0$ and therefore the standard
Euler Gamma function $\Ga(z)$ has no dual symmetry. From (A.9) it
follows that the function $\Ga(z,a,b)\ =\ a\Ga\left(bz,\o {2\pi a}
{b}\right)\left(2\sin\o {\pi a} {b}\right)^{bz-1}$ is symmetrical i.e
$\Ga (z,a,b)\ =\ \Ga(z,b,a)$. Duality exchanges (A.4) with (A.5) and
(A.6) with (A.7).

\beginsection References.

\item[1] J.E.Nelson, T.Regge, Phys.Lett. {\bf B272},(1991)213.
\item[2] J.E.Nelson, T.Regge, F.Zertuche: Nucl.Phys.
{\bf B339},(1990)516: F. Zertuche, Ph.D.Thesis, SISSA (1990),unpublished.
\item[3] V.Moncrief, J.Math.Phys. {\bf 30},(1989)2907.
\item[4] A.Hosoya and K.Nakao, Class.Qu.Grav. {\bf 7},(1990)163.
\item[5] S.Carlip, Phys.Rev. {\bf D42},(1990)2647.
\item[6] S.Carlip and J.E.Nelson, DFTT 67/93 and UCD-93-33, submitted
Phys. Lett. {\bf B}.
\item[7] J.E.Nelson, T.Regge, C.M.P.{\bf 155},(1993)561.
\item[8] See e.g. A.O.Barut and R.Raczka, Theory of group
representations and Applications, World Scientific (Singapore 1986)
p.588.
\item[9] J.E. Nelson and T.Regge, in preparation.
\vfill\eject

Fig.1

\vskip 11truecm

Fig.2.
\bye